\documentclass[]{interact}

\usepackage{epstopdf}
\usepackage[caption=false]{subfig}

\usepackage[numbers,sort&compress]{natbib}
\bibpunct[, ]{[}{]}{,}{n}{,}{,}

\theoremstyle{plain}

\theoremstyle{definition}

\theoremstyle{remark}

\begin{document}


\title{Suppression of magnetic phase transition at high magnetic field and non-Debye's nature of nano-crystalline Gd$_2$CoMnO$_6$: a detail study of physical properties}

\author{
\name{Ilyas Noor Bhatti\textsuperscript{a} \thanks{Ilyas Noor Bhatti. Email: inoorbhatti@gmail.com} Imtiaz Noor Bhatti\textsuperscript{b} Rabindra Nath Mahato\textsuperscript{c}	and M. A. H. Ahsan\textsuperscript{a}	}
\affil{\textsuperscript{a}Department of Physics, Jamia Millia Islamia University, New Delhi - 110025, India;\\
 \textsuperscript{b}Department of School Education, Government of Jammu and Kashmir, India;	\\
\textsuperscript{c}School of Physical Sciences, Jawaharlal Nehru University, New Delhi - 110067, India;}
}

\maketitle
\begin{abstract}
Structural, magnetization, phonon behavior, and dielectric response of nano-crystalline Gd$_2$CoMnO$_6$ have been presented in this paper. The study shows that the material crystallizes in $\textit{P2$_1$/n}$ phase group of monoclinic crystal structure. XPS measurement shows Co$^{2+}$ and Mn$^{4+}$ oxidation states are present in the sample. Magnetization study reveals that sample undergoes a ferromagnetic ordering of Co$^{2+}$ and Mn$^{4+}$ magnetic ions around $T_c$ $\sim$132 K. However we have seen that with the application of external magnetic field the phase transition is largely suppressed. Raman study reveals the presence spin-phonon coupling in Gd$_2$CoMnO$_6$. Dielectric study reveals that the sample shows large dielectric constant and strong dispersion in mid frequency range. The dielectric loss shows there are two relaxation processes present in the material with different relaxation time and which are driven by thermally activated relaxation mechanics. Further, the Nyquist plot and AC conductivity study shows that this sample is non-Debye's in nature.
\end{abstract}


\section{Introduction}
Multiferroicity, magnetoelectric, magnetocapacitance and magnetocaloric properties are seen in rare-earth based double perovskite manganites R$_2$BMnO$_6$ (R = rare-earth and B = Ni/Co) received great attention of researchers due to the potential for device applications.\cite{nsrogado,skumar,gsharma,amtishin,jkmurthy,chagmann,akushino} The superexchange dominated magnetic ordering of (Co/Ni)$^{2+}$ and Mn$^{4+}$ spins give ferromagnetic nature to these compounds where the magnetic phase transition depends on the rear-earth ion which cause structural distortion in double perovskite structure. However the B-site disorder in double perovskite with (Co/Ni)$^{3+}$ and Mn$^{3+}$/B$^{2+}$-B$^{2+}$ result in antiferromagnetic ordering at lower temperature.\cite{goodenough} It is a known fact that the transition temperature $T_c$ of ordered Co/Ni-Mn sublattice decreases with reducing radii of rare-earth cations.\cite{goodenough1,cmeneghini} Apart from this the R (rare earth) ions have the spin moment which shows magnetic interaction via R-R ions and align antiferrimagnetic with B/Mn sublattice at low temperatures.\cite{benitez,kakarla} These compounds have potential to show magnetic field controllable polarization/ capacitance.

There have been several reports on bulk polycrystalline as well as single crystal rare-earth based manganite double perovskite compounds to understand various physical properties. The Ni/Mn based double perovskite has been studied more as compared to Co/Mn. However, there have been several exotic properties have been seen in R$_2$CoMnO$_6$ compounds as well. For instance, recently  La$_2$CoMnO$_6$ a ferromagnetic material with $T_c$ $\sim$225 K \cite{bull}, it is found that material shows spin-phonon coupling in thin-films where both ordered and disordered sample have been investigated.\cite{meyer1} Pr$_2$CoMnO$_6$ is another member of this series which shows a ferromagnetic ordering and feature Griffith's phase in the paramagnetic state close to phase transition.\cite{liu,ilyas1,ilyas2} Various other compounds have been studied and found interesting properties like multiferroicity in (Lu/Y)$_2$CoMnO$_6$,\cite{blasco,sharma} metamagnetic transition in (Eu/Tb)$_2$CoMnO$_6$,\cite{banerjee,moon} and pyroelectricity and magnetocapacitance in Lu$_2$CoMnO$_6$ \cite{vilar}. Among these double perovskites Gd$_2$CoMnO$_6$ have been studied to understand its physical properties whereas most of the efforts have been made to investigate the giant magnetocaloric effect \cite{moon2} at low temperature. Further spin-glass frozen state \cite{xlwang} and metamagnetic transition\cite{marsh} have also been reported for polycrystalline Gd$_2$CoMnO$_6$ close to ferromagnetic transition. Here, our focus is to investigate the detail physical properties of nano-crystalline Gd$_2$CoMnO$_6$. We focus our study to understand the effect of the magnetic field on the magnetic phase transition since earlier literature gives evidence of metamagnetic transition and spin-glass state in this material. Further, low temperature antiferromagnetic ordered Gd$^{3+}$ ions corresponding to Co/Mn sublattice may be effected at the high magnetic field. Dielectric response in rear-earth based manganites are interesting and we will also focus on this study in this paper.

In this paper, we have investigated nano-crystalline Gd$_2$CoMnO$_6$ using magnetization, Raman spectroscopy and dielectric measurements. The sol-gel prepared nano-crystalline Gd$_2$CoMnO$_6$ is characterized by X-ray diffraction for structural study. The sample is found to be single phase and crystallize in  \textit{P2$_1$/n} space group of the monoclinic crystal structure. Magnetization study reveals that the material shows ferromagnetic transition around 130 K and a low temperature antiferromagnetic ordering of Gd sublattice. The magnetic field greatly influences the magnetic structure of Gd$_2$CoMnO$_6$. Raman study shows strong spin-phonon coupling in this material. The dielectric response shows large frequency dispersion, dielectric relaxation is seen in the tangent loss. Impedance spectroscopy shows the material deviates from ideal Debye's model. AC conductivity shows the quantum mechanical tunneling conduction mechanism in this material.

 \begin{figure}[th]
	\centering
		\includegraphics[width=8cm]{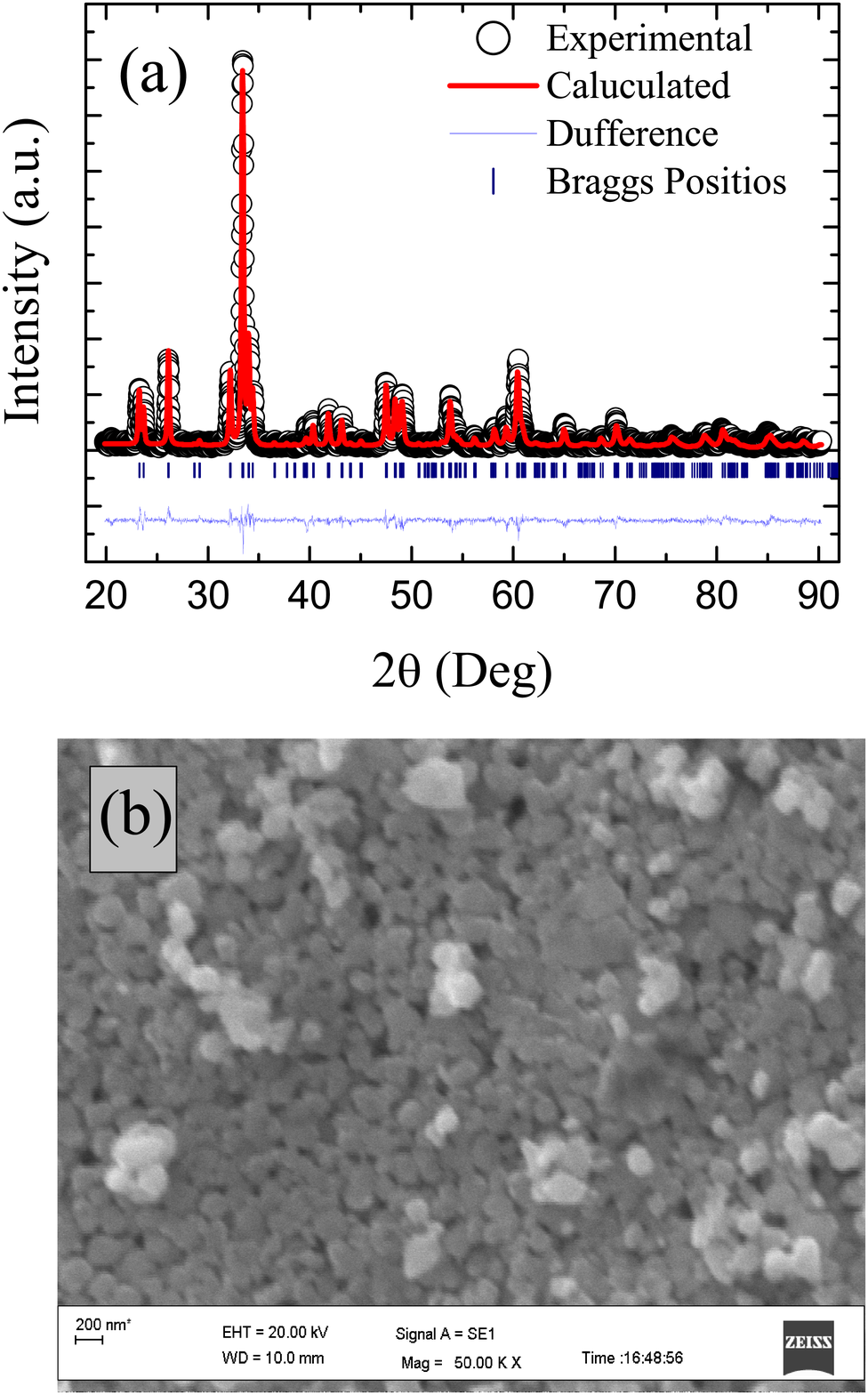}
\caption{(Color online) (a) Rietveld refinement of powder XRD data for nano-crystalline Gd$_2$CoMnO$_6$ (b) SEM micrograph for partical size calculation.}
	\label{fig:Fig1}
\end{figure}

\section{Experimental details}
Nano-crystalline sample of double perovskite Gd$_2$CoMnO$_6$ was prepared by sol-gel method. Stoichiometric amounts of Gd(NO$_3$)$_3$.6H$_2$O and Co(NO$_3$)$_2$.6H$_2$O and C$_4$H$_6$MnO$_4$.4H$_2$O dissolves in doubled distilled water separate beakers. Citric acid is taken n 3:1 ration and dissolved in distilled water. All these solutions are placed on a stirrer for sufficient time to obtain a clear solution. After this, the solutions are mixed in large beaker which is filled up to 400 ml with distilled water. The mixture of solutions was then placed on a magnetic stirrer for 24 hours at 90 $^o$C. After evaporation of water, the solution turned in to gel form which is then heated at 300 $^o$C to obtain solid foam. This foam then collected form the beaker and ground in mortar and pestle to crush it into powder form. The powder sample is then sintered at 850 $^o$C  for 12 hours. Similar method has been adopted for other material \cite{ilyas1,ilyas4,ilyas3}  The sample was then characterization by X-ray diffraction (XRD) (Cu$_{K \alpha}$, $\lambda$ = 1.5406 $\AA$) for structural study. The XRD datat is collected in the 2$\theta$ range of 20-90$^o$ with a step size of 0.02$^o$. The XPS measurements were performed with base pressure in the range of $10^{-10}$ mbar using a commercial electron energy analyzer (Omnicron nanotechnology) and a non-monochromatic Al$_{K \alpha}$ X-ray source (h$\nu$ = 1486.6 eV). The XPSpeakfit software has been used to analyze the XPS data. The magnetic measurements have been carried out on the Physical properties measurement system(PPMS) by Cryogen Inc. Temperature dependent Raman spectra have been collected using Diode based laser (with output wavelength $\lambda$ = 473 nm, coupled with a Labram-HR800 micro-Raman spectrometer. Dielectric measurements in the frequency range from 1 Hz to 5.5 MHz were performed using a computer controlled dielectric setup mounted on CCR with operating temperature range of 15 to 320 K. The dielectric parameters were measured using LCR meter.

\section{Result and Discussions}
\subsection{Structural study}
XRD was done at room temperature for the structural study of the sample. Rietveld refinement was performed with Fullpoof program for the analysis of crystal structure. Fig. 1(a) represents the XRD pattern and Rietveld refinement of the Gd$_2$CoMnO$_6$ sample. Fig. represents black circle for experimental data, red solid line for calculated data, blue lines for difference and navy blue bars represent Braggs positions. The  Rietveld refinement shows that the XRD pattern is well fitted with the calculated one which confirms the phase purity of the sample. The goodness of fit $\chi^2$ and R$_{wp}$/R$_{exp}$ ratio as 1.67 and 1.2487 respectively are acceptable and in agreement with literature.\cite{ilyas3,bhatti1,bhatti2,renu} The crystal structure of Gd$_2$CoMnO$_6$  is monoclinic with space group is \textit{P2$_1$/n}. The lattice parameters obtained from structural analysis $a$, $b$ and $c$ are 5.2755 $\AA$, 5.5616 $\AA$ and 7.5212 $\AA$ respectively. The angle $\beta$ = 90.0747 $^o$ and unit cell volume is 220.6736 $\AA^3$.  Further, to obtain the particle size of the Gd$_2$CoMnO$_6$ sample we have done the scanning electron microscopy. Fig. 1(b) shows the SEM image obtained for the nano-crystalline structure having a uniform distribution of nano-particles. The SEM image is analyzed with ImageJ software. The detail analysis of SEM image shows the average crystalline size for the present compound is $\sim$110.9 nm.       

\begin{figure}[t]
    \centering
        \includegraphics[width=8cm, height=12cm]{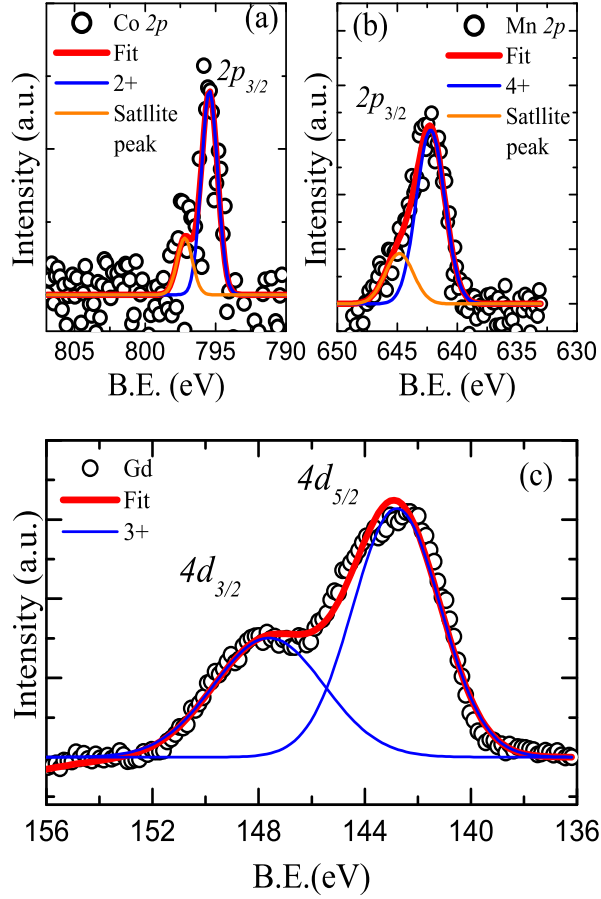}
\caption{(Color online) Figure presents the XPS spectra of (a) Co 2$p$ (b) Mn 2$p$ (c) Gd 4$d$ core levels, where Open circles are experimental data, the red solid lines are the overall fitting and the colored solid lines are the fitted peaks (see figure).}
    \label{fig:Fig2}
\end{figure}

\subsection{X-ray photoelectron spectroscopy (XPS)}
Charge state of cations present in a material decides the electronic and magnetic ground state of that material. Charge state of cations can be understood by XPS measurement on a material here we have recorded the core level spectra of Co, Mn and Gd in Gd$_2$CoMnO$_6$ sample. The XPS spectra of  these elements are fitted with XPS PEAKFIT 4.1. 

Fig. 2a  shows the core level spectrum of Co 2$p$$_{3/2}$ at 796.01 eV corresponding to +2 charge state which is in agreement with reported literature.\cite{wang,qiu,xia,noor1,noor2} Besides the strong Co 2$p$ peak a small less intense satellite peak is also been observed close to Co 2$p$ peaks(see figure).

Fig. 2b shows the Mn 2$p$ core level spectrum along with peak fitting. Here,  Mn 2$p$$_{3/2}$ peaks of core level is presented which is located at 642 eV, which confirms the +4 oxidation state of Mn cations. A relatively small intensity satellite peak is also been seen at higher energy side of Mn 2$p$$_{3/2}$ peak.\cite{ilyas3,noor1,noor2,ida,sachoo,cao}

Fig. 2c shows that the Gd 4$d$ core level spectra, which split into two spin-orbital split peaks Gd 4$d$$_{5/2}$ and Gd 4$d$$_{3/2}$ are located at 142.6 and 148.03 eV respectively (see figure). These peak positions are consistent with literature and corresponds to +3 oxidation state of Gd cations.\cite{reddy,wand} XPS study shows that the oxidation states of Co,  Mn and Gd cations present in Gd$_2$CoMnO$_6$ are +2, +4 and +3 respectively, which is in agreement with the structural results since with Mn$^{+4}$ and Co$^{+2}$ an ordered structure is expected for Mn-Co double perovskite structure.

\subsection{DC Magnetization study}
Temperature dependent DC magnetization $M(T)$ have been measured for Gd$_2$CoMnO$_6$ in both the zero field cooled ($ZFC$) and field cooled ($FC$) protocol. Fig. 3a shows the $M(T)$ data measured in the temperature range from 2 K to 300 K while warming the sample from low temperature. The $M_{ZFC}$ data is collected in $H$ = 100 Oe applied magnetic field ($H$). Whereas $M_{FC}$ is measured at various fields viz, 100 Oe, 250 Oe, 500 Oe and 1000 Oe shown in Fig. 3a. We observe that the $M_{ZFC}$ and $M_{FC}$ remain almost constant with decreasing temperature from 300 K to $\sim$140 K  and both $M_{ZFC}$ and $M_{FC}$ curves overlap on each other.  However, with further decrease temperature the dc magnetization $M$ began to increase sharply. For clarity $M_{ZFC}$ is shown in the inset of Fig. 3a, it is evident that the $M_{ZFC}$ shows a sharp increase in the magnetic moment below 140 K and shows a peak around 130 K whereas with further lowering temperature the moment starts decreasing and attains a stable constant value and forms a plateau. However, at a very low temperature below 40 K, a downfall in $M_{ZFC}$ is observed (see inset Fig. 3a). It is evident from Fig. 3a that the $M_{FC}$ data continuously increases with the decreasing temperature below 140 K. However with further decreasing temperature the $M_{FC}$ data shows a broad peak centered around 50 K and the $M_{FC}$ suffer a downfall in magnetic moment below this temperature down to lowest measurable temperature 2 K. The sharp increase in $M(T)$ data below 140 K is signature of magnetic phase transition from random spin paramagnetic (PM) to ordered ferromagnetic (PM) phase. We observed a large bifurcation between $M_{ZFC}$ and $M_{FC}$ below 130 K.  To better understand the nature of this magnetic phase transition in Gd$_2$CoMnO$_6$ we have measured $M_{FC}$ at various applied magnetic fields (See Fig 3a). We observe that with increasing $H$ the magnetization increases which is expected because more and more spins will orient themselves in the field direction. Further, in the $M_{FC}$ data at the different applied field we observed that with increasing field $T_c$ shifts to the higher temperature. The PM-FM phase transition in Gd$_2$CoMnO$_6$  appears with a sharp increase in magnetization due to cooperative spin interaction. $dM/dT$ vs $T$ plot is often used to estimate the transition temperature $T_c$, the minima i.e. point of inflection in this plot gives $T_c$. Inset of Fig. 3b the shown $dM/dT$ vs $T$ plotted from $M_{FC}$ taken at 100 Oe, it is evident from the figure that $T_c$ for Gd$_2$CoMnO$_6$ is 132 K.

\begin{figure}[t]
    \centering
        \includegraphics[width=8cm]{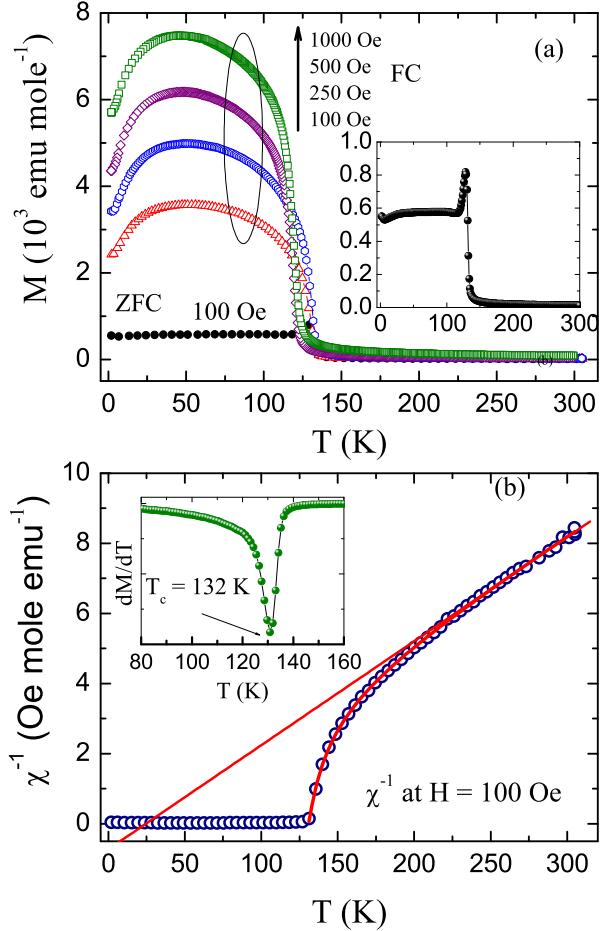}
\caption{(Color online) (a) Temperature dependent magnetization data $M(T)$ shown for Gd$_2$CoMnO$_6$ measured at different low fields. (b) $M(T)$ data plotted in terms of inverse susceptibility $\chi^{-1}$, solid line is fitting due to Curie Weiss Law. Inset shows $dM/dT$ vs $T$ plot showing $T_c$ for Gd$_2$CoMnO$_6$.}
    \label{fig:Fig3}
\end{figure}

 It is evident from the static magnetization data that in Gd$_2$CoMnO$_6$ shows ferromagnetic ordering below 132 K. To understand the magnetic behavior of Gd$_2$CoMnO$_6$ we consider the high spin state for Mn/Co cations. In the above section, we have seen our XPS result confirms that the Mn and Co cations are in +4 and +2 oxidation states respectively. In high spin state the Mn$^{4+}$ (t$_{2g}^{3}$$\uparrow\uparrow\uparrow$ e$_{g}^{0}$$\textendash\thinspace\textendash$) gives S = 3/2 and  Co$^{2+}$ (t$_{2g}^{5}$$\uparrow\downarrow\uparrow\downarrow\uparrow$ e$_{g}^{2}$$\uparrow\uparrow$) yield S = 3/2 with spin only formula.  These cations engage in superexchange interaction below 132 K via Co$^{2+}$-O-Mn$^{4+}$ network and give rise to a ferromagnetic state in Gd$_2$CoMnO$_6$. However, the Gd$^{3+}$ [Xe](4$f^7$) with S = 7/2 present in the material is also magnetic. It is believed that the rare-earth spins become interactive only at low temperatures. The magnetization data (see Fig. 3a) shows a broad peak at 50 K and shows a downfall at low temperature, perhaps it is the anti-ferromagnetic character. In this temperature range the Gd$^{3+}$ spins experience exchange interaction via oxygen as   Gd$^{3+}$-O$^{2+}$-Gd$^{3+}$ and develops an ordered state of Gd sublattice. It is worth to mention that in the temperature dependent magnetization data that the anti-ferromagnetic ordering is observed at low temperature. The antiferromagnetic ordering is attributed because the ordered Gd sublattice is aligned in the opposite direction of the Mn/Co sublattice.

Fig. 3b shows the temperature dependent inverse magnetic susceptibility i.e. $\chi^{^{-1}}$ vs $T$ plotted using $M_{FC}$ data from Fig. 3a at 100 Oe. It is evident from the general feature of the $\chi^{^{-1}}$ plot that above $T_c$  inverse susceptibility shows deviation from conventional Curie Weiss law. It is worth to mention here that for double perovskite compounds such deviation is expected. It is believed that the rare earth (Gd$^{3+}$) sublattice remains non-interactive at high temperature in the network of Co/Mn lattice. Thus a modified Curie Weiss law described as below is used to understand the magnetic behavior in Gd$_2$CoMnO$_6$:\cite{retu,booth}

\begin{eqnarray}
\chi = \frac{C_{TM}}{T - \theta_{TM}} + \frac{C_{Gd}}{T - \theta_{Gd}}
\end{eqnarray}

Here $C_{TM}$ and $\theta_{TM}$  is Curie Constant and paramagnetic Curie temperature for transition metal sublattice. Whereas $C_{RE}$ and $\theta_{RE}$ are the Curie Constant and paramagnetic Curie temperatures for rare earth sublattice. We have performed the non-linear fitting of susceptibility data measured at 100 Oe for  Gd$_2$CoMnO$_6$ shown in Fig. 3b. The fitting is done above $T_c$ in the temperature range 80 K to 300 K. The fitting parameters obtained from fitting of $\chi^{-1}$ with Eq. 1 gives the value of $C_{TM}$ and $\theta_{TM}$ as 3.151(6) emu K mole$^{-1}$ Oe$^{-1}$ and 131 K respectively. The effective paramagnetic moment ($\mu_{eff}^{TM}$) is calculated using these fitting parameters come out to be 5.250(2) $\mu_B/f.u.$, the obtained values are in close to theoretical calculations.  The value of $C_{RE}$ and  $\theta_{RE}$ for Gd sublattice for agreement Gd$_2$CoMnO$_6$ are 32.26 emu K mole$^{-1}$ Oe$^{-1}$ and -18.51(2) K respectively. The negative value of $\theta_{RE}$ signifies that the Gd$^{3+}$ spins are anti-ferromagnetic ordering relative to Mn/Co sublattice. The modified Curie Weiss law in Eq. 1 is presented considering non-interacting Mn/Co and Gd sublattice in this compound. 

\begin{figure}[t]
    \centering
        \includegraphics[width=8cm]{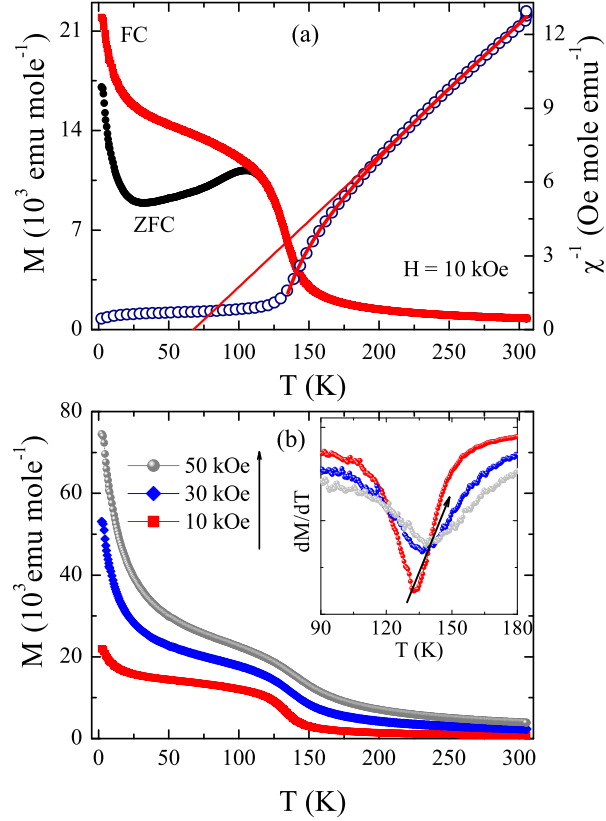}
\caption{(Color online) (a) Temperature dependent magnetization data $M(T)$ shown for Gd$_2$CoMnO$_6$ measured at high fields 10 kOe (left axis), $M(T)$ data plotted in terms of inverse susceptibility $\chi^{-1}$, solid line is fitting due to Curie Weiss Law (right axis). (b) Field Cooled (FC) magnetization data collected at different fields are shown for Gd$_2$CoMnO$_6$. Inset shows $dM/dT$ vs $T$ plot showing $c$ for Gd$_2$CoMnO$_6$.}
    \label{fig:Fig4}
\end{figure}

For further understanding of the low temperature downfall in magnetization data and magnetic transition, we have measured the $M(T)$ at different fields as stated above (see Fig. 3a). It is observed that there is an evolution in the general feature of the $M(T)$ curve. Thus it is quite evident from the magnetization data that the spin structure is greatly influenced at the higher magnetic field. In Fig. 4a we have shown the magnetization data collected at 10 kOe applied field. It is observed that the magnetization data $M_{ZFC}$ and $M_{FC}$ curves show a similar kind of magnetic transition around 130 K. However, below $T_c$ the feature of magnetic data changes greatly as compared to low field data presented in Fig. 3a. Below $T_c$ there is clear bifurcation between the $M_{ZFC}$ and $M_{FC}$ branches. $M_{ZFC}$ decrease with decreasing temperature attains a dip around 30 K and then shown a sharp upturn and increases monotonically till lowest measured temperature. On the other hand the $M_{FC}$ increases with decreasing temperature. This feature in $M(T)$ data measured at the high field of 10 kOe is unlike observed at the low field where both the $M_{ZFC}$ and $M_{FC}$ show an anti-ferromagnetic ordered state. To further deepen our understanding we have measured $M_{FC}$ at the even higher field and presented in Fig. 3b it is quite evident that the low temperature downfall in $M_{FC}$ vanishes. Further, the magnetic transition at 130 K seems to get smoother at high field. Such behavior of spins at the higher field needs to be investigated using some advanced techniques. However, it seems that the spins at the higher field are becoming non-cooperative and result in weakening of magnetic ordering. To further understand this magnetic nature in Gd$_2$CoMnO$_6$ we have plotted the magnetization data in terms of magnetic susceptibility. It is observed that the diversion in $\chi^{-1}$ above $T_c$ decreases which is quite large at low field. The general feature of $M(T)$ curve is largely affected by the increasing magnetic field.  

\begin{figure}[t]
    \centering
        \includegraphics[width=8cm]{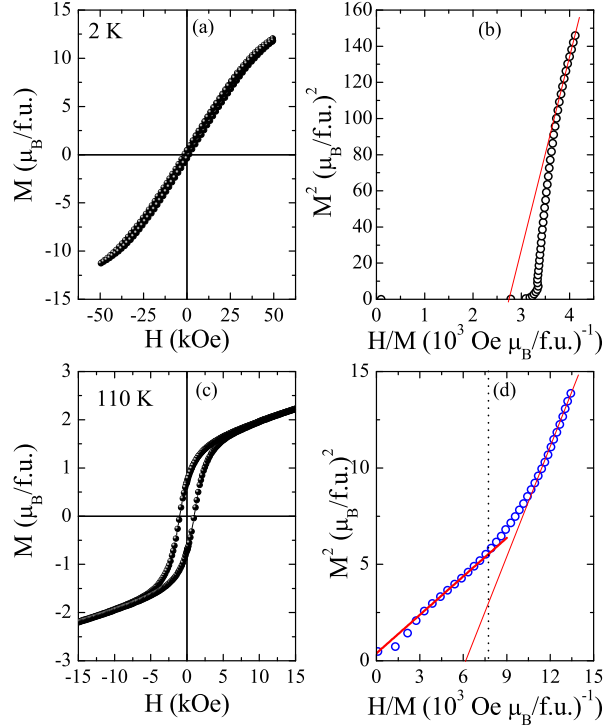}
\caption{(Color online) (a, c) Isothermal data collected at 2 K and 110 K in the applied field of $\pm$ 50 kOe. (b, d) Virgin curves of $M(H)$ (a, c) is plotted in terms of Arrott plot $M^2$ vs $H/M$. }
    \label{fig:Fig5}
\end{figure}

Isothermal magnetization $M(H)$ data measured in both the ferromagnetic state at 110 K and antiferromagnetic state at 2 K up to $\pm$50 kOe applied magnetic field as shown in Fig. 5a and 5b. $M(H)$ curve shows in Fig. 5a shows no hysteresis and the $M(H)$ curve shows a typical anti-ferromagnetic type features. Further, Arrott plot ($M^2$ vs $H/M$) shown in Fig. 5b gives the negative intercept which indicates there is no spontaneous magnetization present in this material at 2 K. However, the $M(H)$ data shown in Fig. 5c is collected in the ferromagnetic ordered state at 110 K seems quite interesting. The hysteresis present in the $M(H)$ curve confirms the ferromagnetic ordering at this temperature. The $M(H)$ data do not saturate even at the highest applied field. The magnetic moment at the highest applied field of 50 kOe is about 3.7 ($\mu$$_B$/f.u.). Whereas the remanent magnetization $M_r$ and coercive force $H_c$ is 0.7 $\mu$$_B$/f.u. and 1000 Oe respectively. In Fig. 5d Arrott plot is shown for $M(H)$ data collected at 110 K exhibits some interesting feature. We observe two slops from the Arrott plot marked two regions, below 15 kOe the extrapolation cut the positive $M^2$ axis and gives a spontaneous magnetic moment $\mu_s$ = 0.6228 $\mu_B$/f.u. However, above this field, a steep increase is observed which when extrapolated intercepts the negative $M^2$ axis which signifies no ferromagnetic ordering present. Thus our results reveals that the magnetic behavior changes with at higher field seen in $M(T)$ is may be due to the non-cooperative behavior of spins at the higher magnetic field in Gd$_2$CoMnO$_6$. We believe that detail local probing and high field investigation is needed to understand this scenario in depth.

\subsection{Temperature dependent Raman study}

Raman spectra have been recorded at different temperatures across magnetic phase transition in the temperature range of 300 K to 10 K Gd$_2$CoMnO$_6$. In Fig. 6a Raman spectra have been shown at selective temperatures for clarity the data each curve is shifted upward. Raman data shows the presence of multiple phonon are active in Gd$_2$CoMnO$_6$ double perovskite material. However, our main focus is on the strongest mode present at 641 cm$^{-1}$ known as breathing mode, since Gd$_2$CoMnO$_6$ crystallizes in \textit{P2$_1$/n} space group thus this breathing mode obeys A$_{1g}$ symmetry. All these modes present in the Raman spectra of  Gd$_2$CoMnO$_6$ are believed to be linked with stretching/bending and rotation of (Co/Mn)O$_6$ octahedra. Symmetric stretching of the (Co/Mn)O$_6$ octahedra gives a strong mode at 636 cm$^{-1}$ and mode at 496 cm$^{-1}$ is due to antisymmetric stretching and bending.\cite{ilive} A weak modes at $\sim$1278 cm$^{-1}$ called second-order overtones of the breathing mode.\cite{meyer} Raman spectra show a little modification in both the intensity and peak positions of each mode with temperature variation. 

To further understand the evolution of Raman spectra with temperature, we have fitted the spectrum collected at different temperatures with the Lorentzian function. Fig. 6b shows the line shape along with Lorentz fitting of A$_1g$ mode at 150 K. From the fitting of Raman spectra we obtained the peak position of Raman mode and line width. The temperature variation of peak position and line width is shown in Fig. 6c and 6d respectively. It is evident, both the position and line width show a deviation from anharmonic behavior around the magnetic ordering temperature $T_c$ 132 K (see magnetization section). The deviation at $T_c$ is known due to the extra scattering mechanism involved due to spin-lattice interaction present in the material. Such deviations in mode frequency shift and line narrowing have been reported for many such systems in earlier literature.

\begin{figure}[t]
    \centering
        \includegraphics[width=8cm]{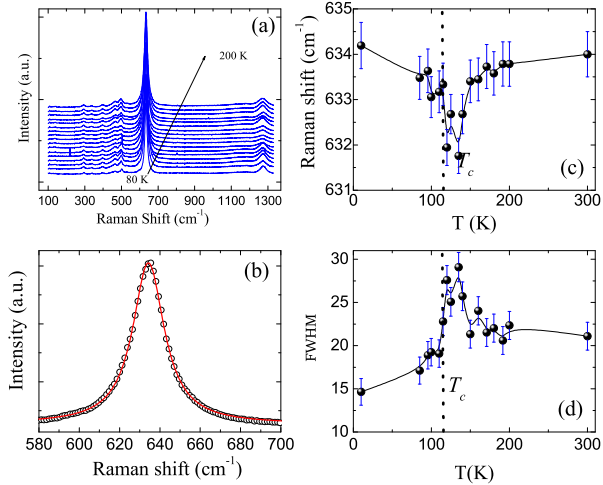}
\caption{(Color online) (a) Raman spectra of Gd$_2$CoMnO$_6$ at selective temperatures.   (b) shows the A$_{1g}$ Raman mode at 641 cm$^{-1}$ solide line is the Lorentzian fit. Temperature Variation of (c) Raman shift (d) FWHM for Raman mode at 641 cm$^{-1}$.}
    \label{fig:Fig6}
\end{figure}

\begin{figure*}[t]
    \centering
        \includegraphics[width=14cm]{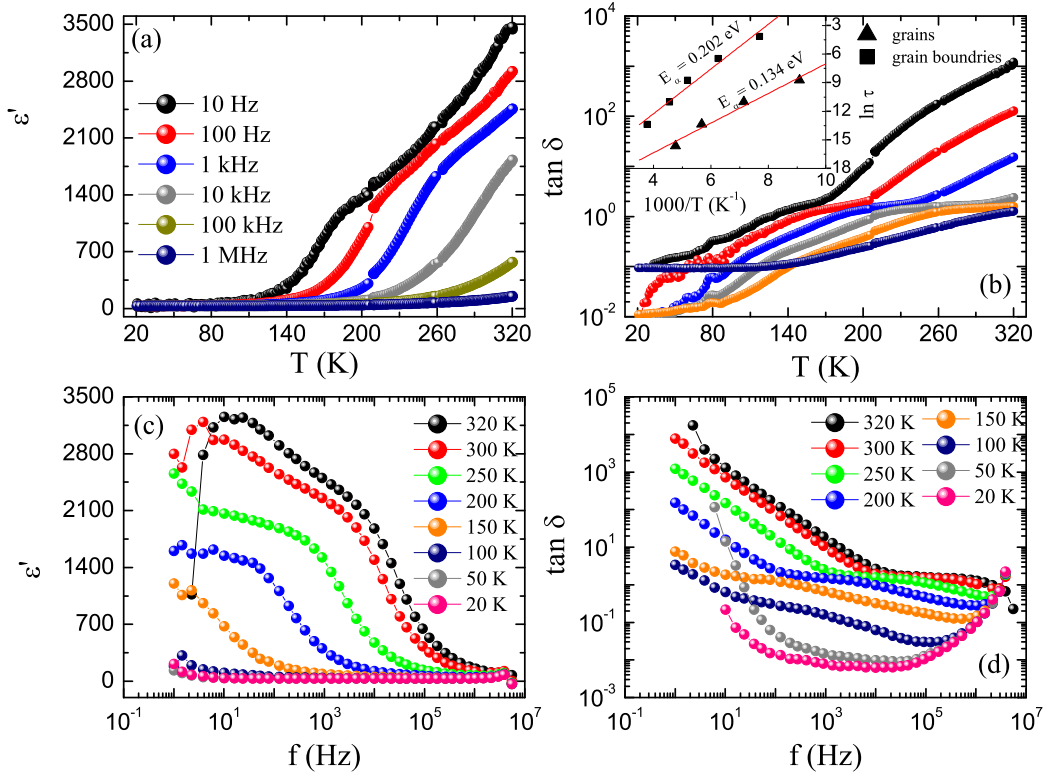}
\caption{(Color online) Temperature dependent (a) dielectric constant ($\epsilon$$^\prime$) (b) loss tangent (tan $\delta$) measure for Gd$_2$CoMnO$_6$. Inset of Fig. (b) plot of relaxation time vs normalized temperature obtained from tangent loss data in Fig. 7(b).  Frequency dependent (c) dielectric constant ($\epsilon$$^\prime$) (d) loss tangent (tan $\delta$)measured at slective temperatures (see figure) in the frequency range 1 Hz to 5.5 MHz.}
    \label{fig:Fig7}
\end{figure*} 

\subsection{Dielectric study}
Dielectric properties of a material are defined in terms of the complex relative permittivity ($\epsilon$), which consists of two components namely a real and an imaginary component. The real part of the complex permittivity is known as the dielectric constant, which defines the measure of energy stored in material from external electric field in which it is placed.  The imaginary part is known as the loss factor. The lag between the change in polarization and the applied electric field makes the permittivity a complex entity  given as:\cite{sch,kao,mansuri,elli,bhatn}
\begin{eqnarray}
\epsilon = \epsilon^{\prime} + i\epsilon^{\prime\prime}
\end{eqnarray}

where, $\epsilon^{\prime}$ $\epsilon^{\prime\prime}$ is the real part and  the imaginary part of the permittivity respectively. The ratio of imaginary part of permittivity to the real part of permittivity is given by following relation and called dielectric loss factor:\cite{sch,kao}

\begin{eqnarray}
tan \delta = \frac{\epsilon^{\prime\prime}}{\epsilon^{\prime}}
\end{eqnarray}

Dielectric constant ($\epsilon^{\prime}$) and loss (tan $\delta$) were measured between 20 K and 300 K temperature at different frequencies for Gd$_2$CoMnO$_6$. Fig. 7a and 7b represent the $\epsilon^{\prime}$ and tan $\delta$ as a function of temperature. Further, the material showing relaxor behavior with various relaxation mechanisms give rise to a plateau in $\epsilon^{\prime}$(T) and respond with peaks in tan $\delta$. For nano-crystalline Gd$_2$CoMnO$_6$  we have observed that at low temperature the $\epsilon^{\prime}$ at all frequencies is merged. However, with increasing temperature, the $\epsilon^{\prime}$ started increasing sharply around 140 K and with further increases in temperature the $\epsilon^{\prime}$ shows strong dispersion with frequency. Further, with increasing frequencies, $\epsilon^{\prime}$ decreases more rapidly. The charge accumulation at the grain boundaries attributes to the higher value of $\epsilon^{\prime}$ at low frequency.By carefully looking into the tangent loss curve tan $\delta$ in Fig. 7b it is found that there are two broad humps at low temperature. Such kind of peak in tan $\delta$ curve is a characteristics of the relaxor phenomenon. With increasing frequency the observed peaks shift towards higher temperature. It is believed that the relaxation mechanism at low temperature is due to grains whereas the one on the higher temperature side is due to grain boundaries.
\begin{figure*}[th]
    \centering
        \includegraphics[width=12cm]{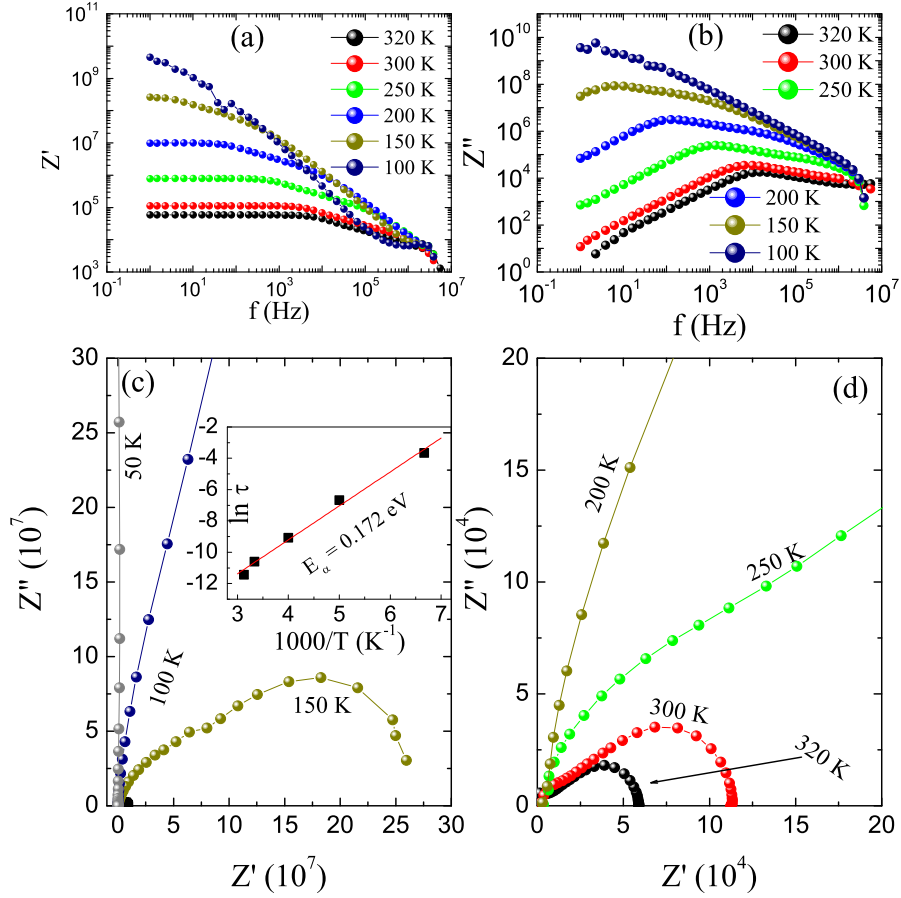}
\caption{(Color online) Frequency dependent data (a) real part of impedance ($Z^{\prime}$) (b) imaginary part of impedance ($Z^{\prime\prime}$ collected at selective temperatures. Inset of Fig (c) shows the variation of relaxation time vs normalized temperature obtained from Fig. 8b. (c), (d) Niqyst plot $Z^{\prime}$ vs $Z^{\prime\prime}$ for Gd$_2$CoMnO$_6$.}
    \label{fig:Fig8}
\end{figure*}

The relaxation mechanism and its origin can be analyzed by the Arrhenius law given by, $\tau_{tan \delta}$ = $\tau_0$ exp(-$E_a$/$K_B$$T$) where, $T$ is the temperature corresponding to the peak observed in the $tan \delta$ curve at a particular frequency $f_{tan \delta}$, $f_0$ is characteristic frequency, $E_\alpha$ is the activation energy of the releaxation process and $k_B$ is the Boltzmann constant. The value of $\tau$ can be obtained from frequency $f_{tan \delta}$ by the relation $\tau$ =1/2$\pi$$f$. Inset of Fig. 7b shows the plot of $\tau$ vs 1000/T for both the relaxation peaks. The data shown in the inset of Fig. 7b obey Arrhenius law and confirm that the relaxation behavior is thermally activated in nature. From the fitting parameters, we have calculated the activation energy for both the relaxation processes as E$_\alpha$ = 0.134 eV for grain and  E$_\alpha$ = 0.202 eV for grain boundaries driven relaxation process. 

The frequency dependent $\epsilon$$^\prime$ and tan $\delta$ over the frequency range 1 Hz to 5.5 MHz for Gd$_2$CoMnO$_6$ at different temperatures is also presented. In Fig. 7c we observed that Gd$_2$CoMnO$_6$ ghave large $\epsilon$$^\prime$ value at low frequencies and high temperatures. The dielectric spectrum is shown in Fig. 7c clearly shows the strong dispersion of dielectric permittivity at low frequencies. Further, the frequency dependent dielectric constant shows that the dispersion creeps towards the higher frequency limit when the temperature of the sample is increased.  In Fig. 7d we present the frequency dependent tangent loss for nano-crystalline Gd$_2$CoMnO$_6$. It is evident from the figure that the tan $\delta$ shows two regions corresponding to what is seen in Fig 7b. The same feature has been observed in the frequency dependent dielectric loss which as we have mentioned earlier is due to the contribution of grains and grain boundaries in the relaxation process.

\subsection{Impedance spectroscopy} 

We have calculated the real and imaginary part of impedance from permittivity data.  Fig. 8 presents the detail impedance data measured for Gd$_2$CoMnO$_6$.  Fig. 8a and 8b shows the $Z^{\prime}$ and  $Z^{\prime\prime}$  data plotted against frequency, data is shown on logarithmic scale for clarity. Both $Z^{\prime }$ and  $Z^{\prime\prime}$ shows similar trend with temperature evident from figures that both parameters decreases at high temperature.  $Z^{\prime\prime}$  data id of particular interest since we can identify the peak in the to get the activation energy of relaxation process involved. In ($Z^{\prime\prime}$) curves there emerges a peak which reaches a maximum value at  $Z^{\prime\prime}_{max}$ corresponding to each curve, and interestingly $Z^{\prime\prime}_{max}$  creeps to high frequency limit on increasing temperature.  It is this peak shift which suggests relaxation time constant would reduce with increasing temperature. 
The most probable relaxation time ($\tau$) of relaxation system  can be estimated by identification the $Z^{\prime\prime}_{max}$  in the $Z^{\prime\prime}$  vs log ($f$) plots following  given relation:\cite{elli,bhatn}

\begin{eqnarray}
\tau = \frac{1}{\omega} = \frac{1}{2 \pi f}
\end{eqnarray}

where $\tau$ and $f$ is the is relaxation time and frequency respectively. 
Relaxation behavior can further be comprehend by plotting relaxation time $\tau$ vs 10$^3$/$T$ (K$^{-1}$)and analyzing its behavior with temperature. Inset of Fig. 8c shows the temperature variation of $\tau$ and it is found to follow the Arrhenius behavior given as:\cite{elli,bhatn,ali}
 
\begin{eqnarray}
\tau_b = \tau_0 exp \left( \frac{-E_a}{k_BT} \right)
\end{eqnarray}

where $\tau_0$ is the pre-exponential factor, k$_B$ the Boltzmann constant and $T$ the absolute temperature. The data in Fig. 8c is well fitted with Eq. 5 and the activation energy (E$_a$) thus calculated from fitting data and its value is 0.172 eV for this material.

Fig. 8c and 8d represent Nyquist plots ($Z^{\prime}$ vs $Z^{\prime\prime}$) for Gd$_2$CoMnO$_6$. Fig. 8c-d reveal that the plot gives the combination of two semicircles in the full temperature range and the radius of the first semicircle decreases with increasing frequency. The first semicircle is due to grains and the second semicircle is due to the grain boundaries. Results from the dielectric loss and impedance spectroscopy show that the relaxation has two contributions grains and grain boundaries.   

\begin{figure}[th]
    \centering
        \includegraphics[width=8cm]{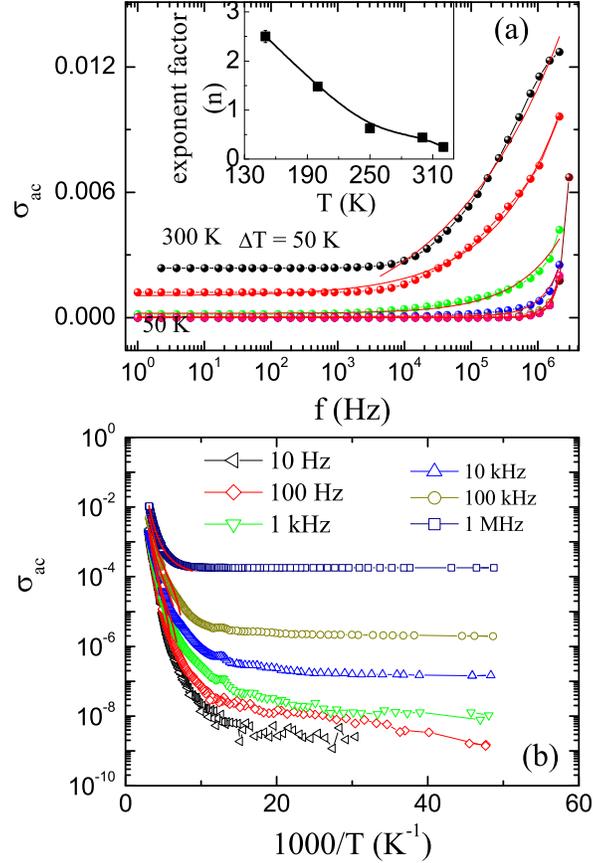}
\caption{(Color online) (a) Presents AC conductivity ($\sigma_{ac}$) as a function of frequency. Solid red lines are fitting due to Eq. 7. Inset of Fig.(a) shows the exponent $n$ versus the temperature for non-crystalline Gd$_2$CoMnO$_6$. (b) The variation of $\sigma_{ac}$ with absolute temperature (10$^3$/T) is shown for Gd$_2$CoMnO$_6$, solid lines are fitting due to Eq. 8.}
    \label{fig:Fig9}
\end{figure}

\subsection{Electric conductivity}

To shed some light on charge hoping and transport mechanism, AC conductivity in Gd$_2$CoMnO$_6$ is studied which is calculated by following relation:\cite{sing}

\begin{eqnarray}
\sigma_{ac} = \epsilon_0 \omega \epsilon^{\prime\prime}. 
\end{eqnarray}
where, $\epsilon_0$ is free space permittivity, $\sigma_{ac}$ is AC conductivity, $\omega$ is angular frequency and $\epsilon^{\prime\prime}$ is imaginary part of permittivity.

Frequency dependent AC conductivity i.e. $\sigma_{ac}$ vs $f$ measured at different temperatures is shown in Fig. 9a for Gd$_2$CoMnO$_6$, which reveals that at low frequencies the conductivity shows plateau region thus remains frequency independent. The dc conductivity ($\sigma_{dc}$) mainly contributes to the conductance process in this region of frequency. At higher frequencies, the conductivity does not show any plateau like behavior as seen at low frequencies, however it increases sharply with increasing frequency. Further, with increasing temperature the plateau seems to shifts to higher frequency with increasing temperature. The frequency independent region gives no signature of hopping charge carriers at low frequencies. High frequency  AC conductivity obey Jonscher's Universal Power Law given as:\cite{thakur}

\begin{eqnarray}
\sigma_{ac} = \sigma_{dc} + A\omega^n
\end{eqnarray}

when A is a temperature dependent constant and gives the degree of polarizability, $\omega$ = 2$\pi$$f$ and $0 \leq n \geq$ is an temperature dependent exponent. The mobile ions and lattice interaction usually described by the value of power law exponent $n$.

The frequency dependent $\sigma_{ac}$  is fitted with the power law in complete  frequency range as shown in Fig. 9a. From the fitting parameter, $n$ is obtained for all temperatures and plotted in the inset of Fig. 9a. It is observed that the exponent $n$ is less than 1 at room temperature and increases with decreasing temperature which shows that the system is non-Debye's in nature. 

Fig. 9b presents the $\sigma_{ac}$ vs 10$^3$/T plot for Gd$_2$CoMn$_6$ selective frequencies, which shows that with increasing frequency the conductivity increases. The temperature dependent behavior of AC conductivity can be understood by fitting the data in Fig. 9b with equation below:\cite{kim,puli,chen}

\begin{eqnarray}
\sigma_{ac} = \sigma_{0}exp\left(\frac{-E_a}{k_BT}\right)
\end{eqnarray}
where $\sigma_{0}$ is pre-exponent factor, $k_B$ is Boltzmann constant and $E_a$ is the activation energy.
From the fitting parameters we have calculated that activation energy and observed that the activation energy decreases with increasing frequency. The activation energy obtained for different frequencies are given as E$_a$(10 Hz) = 3.24 eV, E$_a$(100 Hz) = 3.09 eV, E$_a$(1 kHz) = 3.00 eV, E$_a$(10 kHz) = 2.27 eV, E$_a$(100 kHz) = 2.18 eV and E$_a$(1 MHz) = 2.06 eV

\section{Conclusion}
Gd$_2$CoMnO$_6$ nano-crystalline sample was successfully synthesis by the sol-gel method and its structural, magnetic and dielectric properties were studied. Structural study shows that the sample crystalline \textit{P2$_1$/n} space group of monoclinic structure with average crystallite size $\sim$110.9 nm. Magnetization study shows that the sample develops a ferromagnetic ordered state below 131 K. Further, it is found that the magnetic field greatly influences the magnetic properties and tune the phase transition. It is observed that the sample becomes almost-paramagnetic like at a high applied field of 50 kOe. Isothermal magnetization study confirms that low field sample behaves like ferromagnetic material with a finite value of the spontaneous magnetic moment. At high field limit, the spontaneous moment was not found from the Arrott plot which indicates the absence of magnetic ordering. Raman study shows the strong spin-phonon coupling present in Gd$_2$CoMnO$_6$. Dielectric study shows that the relaxation behavior is thermally activated process and contribution come from both grains and grain boundaries in the dielectric loss. The relaxation time is distributive in nature and thus the material is non-Debye. The impedance spectroscopy analysis and Nyquist plot study confirm that the grain and grain boundaries. The Nyquist plot deviates from ideal semicircles and further confirm the non-Debye's behavior of this material. The conductance behavior is strongly frequency dependent and the value of exponent factor $n$ suggests that Gd$_2$CoMnO$_6$ is non-Debye in behavior.

\section*{Acknowledgment}
We acknowledge AIRF (JNU), New Delhi India, MNIT Jaipur, India and UGC-DAE-Consortium Indore for magnetic measurement facilities, XPS measurement and Raman data respectively. We thanks Dr. A. K. Pramanik for dielectric measurement and also acknowledge UPEA-II funding for LCR meter. Ilyas Noor Bhatti acknowledges University Grants Commission, India for financial support. 

\section*{Disclosure statement}

The authors declare that they have no conflict of interest.

\end{document}